# Gettering in polySi/SiO$_x$ passivating contacts enables Si-based tandem solar cells with high thermal and contamination resilience

Alireza Assar[1*ξ], Filipe Martinho[2*ξ], Jes Larsen[3], Nishant Saini[3], Denver Shearer[1], Marcos V. Moro[4], Fredrik Stulen[5], Sigbjørn Grini[5], Sara Engberg[2], Eugen Stamate[1], Jørgen Schou[2], Lasse Vines[5], Stela Canulescu[2*], Charlotte Platzer-Björkman[3], Ole Hansen[1*]

[1] DTU Nanolab, National Centre for Nanofabrication and Characterization, Technical University of Denmark, DK-2800, Kongens Lyngby, Denmark

[2] Department of Photonics Engineering, Technical University of Denmark, DK-4000 Roskilde, Denmark

[3] Division of Solar Cell Technology, Department of Materials Science and Engineering, Uppsala University, 75236 Uppsala, Sweden

[4] Department of Physics and Astronomy, Uppsala University, Box 516, 75120 Uppsala, Sweden

[5] Department of Physics, University of Oslo, 0371 Oslo, Norway

*Corresponding authors: Alireza Assar (alhaj@dtu.dk), Filipe Martinho (filim@fotonik.dtu.dk), Ole Hansen (ohan@dtu.dk), Stela Canulescu (stec@fotonik.dtu.dk)

ξ These authors contributed equally.

**Abstract**

Multijunction solar cells in a tandem configuration could further lower the costs of electricity if crystalline Si (c-Si) is used as bottom cell. However, for direct monolithic integration on c-Si, only a restricted number of top and bottom cell architectures are compatible, due to either epitaxy or high temperature constraints, where the interface between subcells is subject to a trade-off between transmittance, electrical interconnection, and bottom cell degradation. Using polySi/SiO$_x$ passivating contacts for Si, this degradation can be largely circumvented by tuning the polySi/SiO$_x$ stacks to promote gettering of contaminants admitted into the Si bottom cell during the top cell synthesis. Applying this concept to the low-cost top cell chalcogenides Cu$_2$ZnSnS$_4$ (CZTS), CuGaSe$_2$ (CGSe) and AgInGaSe$_2$ (AIGSe), fabricated under harsh S or Se atmospheres above 550 °C, we show that increasing the heavily-doped polySi layer thickness from 40 to up to 400 nm prevents a reduction in Si carrier lifetime by one order of magnitude, with final lifetimes above 500 µs uniformly across areas up to 20 cm$^2$. In all cases, the increased resilience was correlated with a 99.9% reduction in contaminant concentration in the c-Si bulk, provided by the thick polySi layer, which acts as a buried gettering layer in the tandem structure without compromising the Si passivation quality. The Si resilience decreased as AIGSe > CGSe > CZTS, in accordance with the measured Cu contamination profiles and higher annealing temperatures. An efficiency of up to 7% was achieved for a CZTS/Si tandem, where the Si bottom cell is no longer the limiting factor.

# 1 Introduction

Photovoltaics (PV) can now provide the lowest Levelized Cost of Energy (LCOE) for more than two-thirds of the world's population [1]. According to industry surveys conducted in 2020, the largest fraction of this LCOE corresponds to the Balance of System (BOS) costs, which represent all the costs not related to the actual PV module. BOS now comprises over 66% of the total cost of a PV system, compared to 34% for the PV module, and this proportion is expected to remain constant towards the next decade, even as the total cost of a PV system continues to decrease [2]. Moreover, of this 34% fraction from the PV module, nearly half comes from materials and consumables unrelated to the cell fabrication, used in standardized module assembly (whether it be wafer-based or monolithic thin film modules) [2]. With these numbers, the PV industry is now at a stage where 80% of the costs are independent of the core solar cell technology used, which means that aiming for low-cost cell types and fabrication methods alone is unlikely to have a significant impact on the overall LCOE of PV. Instead, since the majority of the BOS costs scale with the total area of the PV system (in $m^2$), the cost of PV (e.g., in \$/kWh) is now mostly dependent on the areal energy density of the PV system ($kWh/m^2$). For that reason, the PV industry has been recently shifting to concepts that maximize the $kWh/m^2$ energy density of a PV system, leading to a notable increase in systems with axis tracking, bifacial or half-cut cells, high-efficiency cell concepts such as n-type monocrystalline Si (mono-Si) with passivated contacts (tunnel oxide passivated contacts, TOPCon, or silicon heterojunction, SHJ) [3] and, in particular, the development of tandem cell concepts, which are expected as early as 2023 by the industry [2]. Of these, tandem cells may potentially enable the largest increase in the $kWh/m^2$ output of a PV module, as it is the only feasible method that can exceed the single-junction Shockley-Queisser efficiency limit [4], or the Auger limit of 29.4% for Si solar cells [5].

With a 93% market share estimated in 2020 [2], crystalline Si (c-Si) is in a leading position to transition the industry to double-junction (2J, or tandem) cell architectures. The Si bandgap of 1.12 eV is close to the ideal value for a bottom cell (0.9-1.0 eV [6]), and mono-Si cells have the potential to achieve the lowest LCOE of any tandem configuration [7,8]. However, there is still only a limited set of top cell materials known to be compatible with c-Si, and only a restricted number of cell and module configurations is considered feasible or industrially attractive, as reviewed recently [7,9]. In particular, the direct monolithic tandem approach – *i.e.*, on-wafer integration without additional substrates or adhesives, usually resulting in two-terminal (2T) devices – is regarded as the most scalable and compatible with current high-throughput industrial tools. However, so far only two direct monolithic approaches have demonstrated AM 1.5G tandem efficiencies exceeding 20%: 1) perovskite/Si tandems, which have achieved an efficiency of 29.5% certified in 2020 [10], and 2) epitaxial GaAsP/Si tandems based on GaP/Si templates on Si(100) substrates, with a 25.0% efficiency also reported in 2020 [11]. In the first case, the long-term stability of organometal halide perovskites remains an open question [12]. In particular, a reduced longevity can severely impact the LCOE, when compared to the 25 and 30 year warranty at > 92% performance currently achieved by new Si and CdTe modules, respectively [13,14]. In the second case, the tandem efficiency still lags behind single-junction Si and the total cell costs increases considerably due to the processing of III-V materials [15]. Chalcogenide compounds such as Cu(In,Ga)(S,Se)$_2$ (CIGSSe) and Cu$_2$ZnSn(S,Se)$_4$ (CZTSSe) could be suitable alternative top cell materials for a chalcogenide/Si tandem, as they combine the potential of low-cost synthesis, low toxicity, high stability and a bandgap tunability by means of S-Se variations or cationic substitutions [16–18]. While certain low-bandgap chalcogenides (1.0-1.2 eV) have achieved > 20% efficiencies (23.35% for a Cs-treated, Ga-graded CIGSSe absorber [19]), high-bandgap chalcogenides (> 1.5 eV) have historically lagged behind in terms of efficiency, due to a comparatively less favorable mix of higher concentration of deep bulk defects, detrimental grain boundaries and interface recombination [16,17,20,21]. However, several types of high-bandgap chalcogenides have seen notable advances recently. The highest efficiency currently stands at 15.5% for CuInGaS$_2$ (CIGSu, 1.55-1.60 eV), reported in 2016 [22], with an unpublished result of 17% in 2017 [23], and a 15.2% device demonstrated in 2021 from a non-toxic process (Cd-free, no H$_2$S) [24]. Other examples are CuGaSe$_2$ (CGSe, 1.65-1.70 eV) which achieved 11.9% in 2017 [25], and Ag(In,Ga)Se$_2$ (AIGSe, 1.70-1.75 eV), with a 9.4% cell reported in 2017 [26]. Other wide-bandgap chalcopyrite and kesterite compounds are currently below the 10% efficiency level, but all have received less attention than their low-bandgap counterparts due to the relatively recent interest in Si-based or thin-film-based tandem concepts [27]. To combine any of these wide-bandgap chalcogenides with Si in a direct monolithic approach, the Si bottom cell must retain its quality after the top cell fabrication, which includes resistance to contamination and resilience to high temperature steps (> 500 °C), typical for chalcogenide compounds. For mono-Si cells, this implies a post-processing effective carrier lifetime ($\tau_{eff}$) above 1 ms, or an implied open-circuit voltage (i-$V_{oc}$) above 700 mV. Recently, we have proposed that this can be achieved using a double-sided TOPCon Si structure for the bottom cell, as its polycrystalline Si on oxide (polySi/SiO$_x$ or POLO) passivating contacts combine the potential for high efficiency and thermal stability. Due their excellent carrier selectivity and interface passivation quality [28], TOPCon Si cells have achieved up to 26.1% efficiency [29] and their industrial use is quickly expanding [2,30]. Moreover, we have shown that the bottom cell degradation can be minimized with a TiN-based diffusion barrier layer up to 10 nm thick [31,32]. However, while $\tau_{eff}$ > 1 ms could be demonstrated for Si post-processing, our results revealed a clear trade-off between bottom cell protection, electrical interconnection between subcells, and optical transmittance of the diffusion barrier layer. This fundamental tandem trade-off triangle for chalcogenide/Si tandem cells is illustrated in **Figure 1**, based on a summary of our previous findings [31,32]. These three requirements are essential in any monolithic tandem configuration, and although they appear in different forms depending on the combination of materials and the tandem architecture chosen, they fundamentally relate to the subcell interconnection (or recombination) layer(s). The subcell interconnection is therefore subject to a set of constraints. Most importantly, it must provide a recombination path to the inner carriers with minimal voltage losses, which can be achieved by either a tunnel junction, a carrier selective membrane or simply a conductive layer with appropriate band edge positions. This requirement must be achieved with a minimal loss in transparency for the light that reaches the Si bottom cell (above 720 nm or below 1.72 eV for a bandgap-matched tandem). Then, for each configuration, other accessory characteristics could be useful. For a chalcogenide/Si tandem, its success is contingent on the interconnection layers(s) preventing the diffusion of contaminants into the Si bulk, especially copper, which in Si exhibits a variety of defect complexes and copper silicide precipitation, most of which leading to high non-radiative recombination [33]. In some cases, protection against solvent or sputtering damage to the bottom cell may be useful. In other cases, there are constraints on the lattice parameters if lattice-matched epitaxy is required. Moreover, anisotropic conductivity, with high out-of-plane conductivity and low in-plane conductivity, could be useful to mitigate shunting paths between top and bottom cells [34] or between module segments in monolithic tandem modules (*i.e.* thin film modules, not wafer-based) [35]. Finally, the subcell interconnection layer(s) should ensure a good mechanical stability and adhesion between adjacent layers. For a more complete review on this topic, the reader is referred to ref. [7].

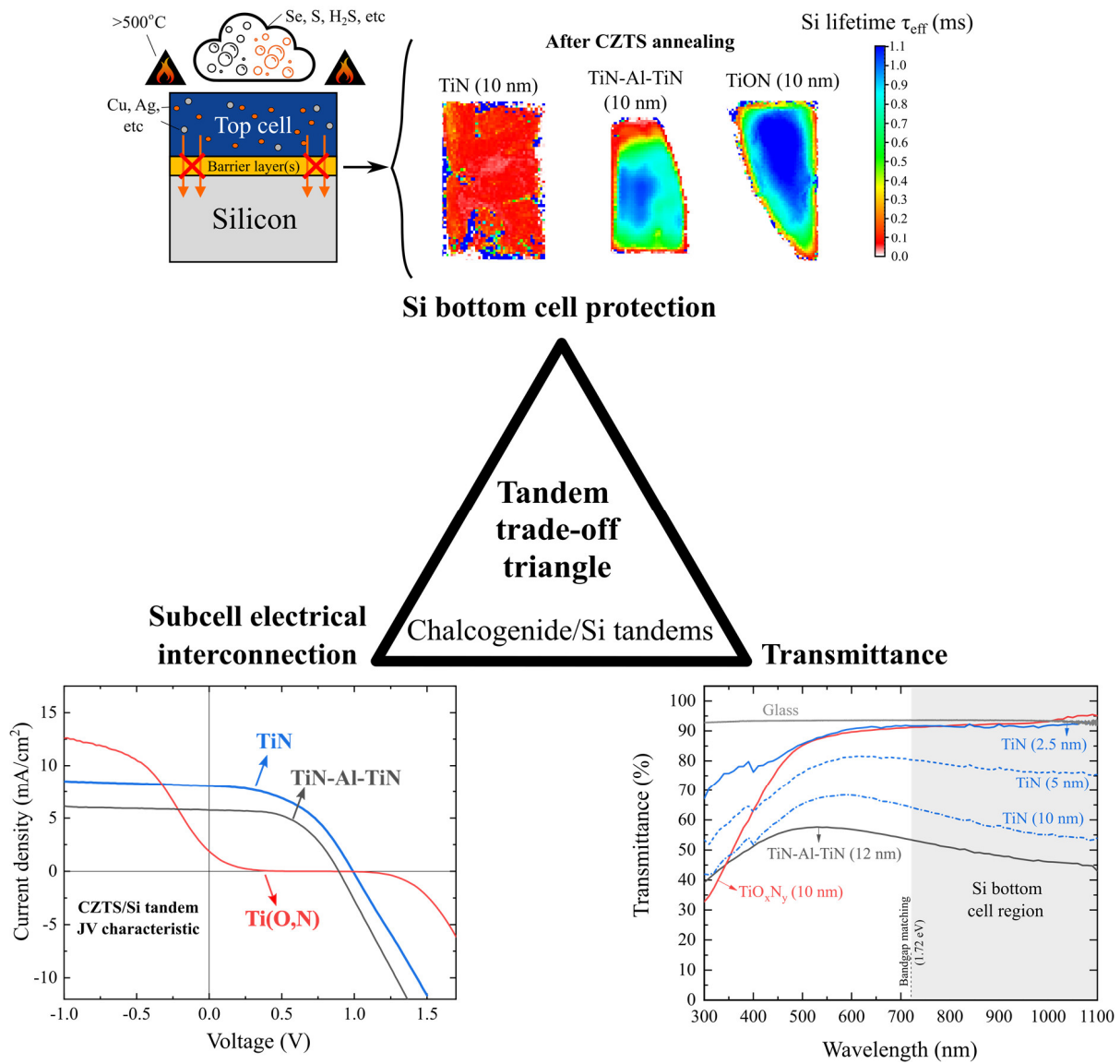

**Figure 1** – The tandem trade-off triangle for a direct monolithic chalcogenide/Si tandem, showing the balance between Si bottom cell protection (top), subcell electrical interconnection (left) and barrier layer transmittance (right). Figures adapted from previous work with permission [32], copyright © 2020 American Chemical Society.

The best device reported in our previous work, based on a $Cu_2ZnSnS_4$/Si tandem with 3.9% efficiency, was still severely limited by an insufficient protection of the Si bottom cell [32]. Since the emphasis must be on high efficiency, the feasibility of direct monolithic chalcogenide/Si tandems is therefore critically dependent on solving this trade-off directly on high-efficiency Si structures. For instance, in other works, chalcogenide/Si tandems have been developed using a transparent conductive oxide (TCO), such as ITO [36] or FTO and variations thereof [37]. However, in these studies the Si bottom cells were based on the Aluminum back surface field (Al-BSF) structure, which is intrinsically limited to < 22% efficiency due to the lower passivation quality of the rear interface [38]. In this work, we report a significant improvement in the resolution of this tandem triangle trade-off, by combining our previously reported TiN-based diffusion barrier with a modification of the polySi/$SiO_x$ passivating contacts of our TOPCon Si bottom cell. We show that both the n+ and p+ polySi layers can be tuned to act as additional barriers by promoting a gettering of the contaminants introduced during the high-temperature step of the top cell growth. Previously, it has been reported that polySi/$SiO_x$ contacts can act as in-diffusion barriers, protecting the mono-Si base wafer during the doping process, both for *in-situ* doping during the low-pressure chemical vapor deposition step (LPCVD, as used in this work) or for *ex-situ* diffusion doping [39]. Additionally, under certain deposition conditions, it has been shown that heavily phosphorous-doped and boron-doped polySi layers can getter pre-existing contaminants from the mono-Si base. For a sufficiently thin silicon oxide (<1.3 nm), the polySi gettering efficiency is comparable to that of $POCl_3$- and $BBr_3$-diffused Si (*i.e.*, without PolySi, conventional phosphorous and boron diffusion gettering, PDG and BDG, respectively) [40,41]. However, polySi/$SiO_x$ contacts achieve this gettering without compromising the quality of the interface passivation, leading to an overall improvement in the Si carrier lifetime $\tau_{eff}$ without the need for extra removal steps, such as etching of the gettering layer(s) [40,42]. This property is of particular interest for a tandem application, because at least one of the polySi layers is effectively buried in the tandem structure, and therefore cannot be removed after the processing. In the abovementioned studies, the key parameters to control the polySi barrier and gettering effects were found to be the polySi deposition method, thickness, doping density, doping profile and annealing temperature. In this work, we make use of these effects by tuning the thickness and doping profile of the n+ and p+ polySi layers during the LPCVD step, and find a clear improvement in the bottom cell resilience. In turn, this increased resilience allows a reduction of the TiN barrier layer down to 2.5 nm, achieving a near-ideal transmittance of above 90%. In contrast to single-junction Si TOPCon devices (1J TOPCon), where every 10 nm of polySi implies approximately a 0.5 mA/$cm^2$ loss in current density due to UV parasitic absorption [43], the increase in polySi thickness in this tandem concept does not reduce the current, since the high-energy photons are absorbed in the top cell. This allows us to increase the polySi thickness from 40 nm, typical in 1J TOPCon Si cells, to up to 400 nm. We find evidence of impurity gettering with increasing polySi thickness, in particular for Cu, and find that both 200 nm and 400 nm are satisfactory to achieve minimal bottom cell degradation, with a uniform $\tau_{eff}$ well above 500 μs across an area of several $cm^2$ of tandem samples. Based on these results, we tested our proposed Si bottom cell platform in a cross-laboratory collaboration for the study of different top cells, including CZTS, CGSe and AIGSe. For a visualization of the corresponding bottom cell TOPCon Si structures, chalcogenide top cell configurations and the subcell interconnection layer based on TiN, the reader is referred to **Figure 2** below.

In all cases, we observe that the Si bottom cell endures the top cell fabrication with minimal degradation, highlighting the general applicability of our proposed TOPCon Si architecture for advancing chalcogenide/Si tandems. For CZTS/Si, we achieved a tandem efficiency of 6.8% ($V_{oc}$ = 1.08 V), the highest of its kind. For CGSe/Si, a cell with 5.1% efficiency and $V_{oc}$ of 1.3 V was obtained. This improvement to the tandem trade-off triangle problem provides a clear roadmap for a >10% chalcogenide/Si tandem in the near-term, simply based on optimizations of Si backend processing and chalcogenide deposition. Furthermore, we surmise that a >20% chalcogenide/Si tandem could be possible in the medium term, contingent only upon continuous efficiency advances for wide-bandgap chalcogenides.

## 2 Experimental Section

The fabrication of the polySi/SiO$_x$ passivating contacts, SiN:H hydrogenation and TiN process flow follows recipes developed in previous work [31,32]. The n+polySi layers, fabricated using low pressure chemical vapor deposition (LPCVD), were deposited sequentially. First, standard 40 nm n+polySi and p+polySi layers were fabricated using the previously reported recipe, and the Si carrier lifetime was measured (without hydrogenation). Then, 160 nm n+polySi was deposited to achieve 200 nm, and a further 200 nm was deposited in order to achieve n+polySi layers with 400 nm thickness. Before each sequential polySi deposition, the wafers were RCA cleaned, followed by an HF dip shortly before each new deposition to remove the native oxide. After the fabrication of the thick n+polySi layers, the samples were annealed at 850 °C for 15 min, followed by a SiN:H hydrogenation recipe to improve the interface passivation quality. The front SiN:H was removed in buffered HF in order to fabricate the TiN layer and the subsequent top cell layers, but the rear SiN:H was kept throughout the entire top cell fabrication to protect the backside. The rear SiN:H was removed only before the deposition of the back electrode, which used 500 nm of thermally evaporated Ag or Al.

The effective minority carrier lifetime ($\tau_{eff}$) in Si was mapped down to a minimum resolution of 0.5 mm$^2$ using the microwave-detected photoconductance decay method (μ-PCD), in a steady-state configuration at 1 sun illumination (~ $10^{15}$ cm$^{-3}$ injection), using an MDPmap lifetime scanner from Freiberg Instruments. The median $\tau_{eff}$ values obtained from the mappings were cross-checked using a Sinton lifetime tester WCT-120. Depending on the lifetime of the samples, two modes were used. For $\tau_{eff}$ > 200 μs, a transient mode (tr-PCD) was chosen, using a flash lamp with a time constant of 30 μs. For $\tau_{eff}$ < 200 μs, a quasi-steady-state mode (QSSPCD) was used with a 2.1 ms time constant for the light source. The $\tau_{eff}$ values from the WCT-120 tool correspond to an average over a large central portion of the wafer, and were generally found to be comparable to MDPmap values, unless otherwise specified. In general, the lifetime was measured from the backside (SiN:H/p+polySi), before and after the top cell fabrication.

The top cell chalcogenides fabricated in this work were $Cu_2ZnSnS_4$ (CZTS), $CuGaSe_2$ (CGSe) and $AgInGaSe_2$ (AIGSe). Their fabrication route, annealing parameters and corresponding bottom cell configuration for the tandem experiments are summarized in **Table 1**. For CZTS, two different recipes (CZTS-D and CZTS-U), developed in two different laboratories, were used for tandem integration and testing of the bottom cell resilience. Both recipes are based on a two-step process, where a Cu-Zn-Sn-S precursor is deposited by cosputtering, followed by an annealing step in the presence of excess sulfur. For the CZTS-D process, the precursor was deposited by cosputtering from Cu, SnS and ZnS targets under an Ar atmosphere, without any intentional substrate heating. The annealing step was done by placing the samples in a graphite box and loading them in a quartz tube furnace. In the box, a total of 50 mg of S pellets and 5 mg of Sn powder were introduced. The annealing profile was 575 °C for 45 min, with a heating rate of 20 °C/min starting from room temperature. A temperature of 560 °C was also used in separate experiments, to test the Si bottom cell resilience. An initial $N_2$ pressure of 175 mbar is introduced in the tube, which expands to 310 mbar during the high-temperature dwelling step. Further details on the cosputtering, annealing setups and top cell fabrication steps can be found in previous work [44]. In the CZTS-U process, the precursor was fabricated from cosputtering of CuS, SnS and ZnS targets under an Ar atmosphere, with the substrate kept at 250 °C. For the annealing step, a tube furnace was used, comprised of three zones: a cold load-lock at T~40 °C, a cooling zone (equipped with an external Cu coil), and the main hot zone. The samples were loaded into a graphite box with 80 mg of S. Prior to the process, the hot zone was stabilized at the nominal annealing temperature of 582 °C, with an Ar pressure of 350 mbar. The annealing was then started by loading the box from the cold loadlock to the hot zone, using dwelling times of 1 min and 13 min. After the designated dwelling, the samples were instantly pulled into the cooling zone, where they cool down to 100 °C in approximately 10 min. Further details on the cosputtering and annealing setups can be consulted in previous work [45,46]. For both CZTS processes, the CZTS thickness was controlled by the cosputtering time, resulting in CZTS absorbers with 300-350 nm after annealing. The CGSe absorbers were fabricated by coevaporation from Cu, Ga and Se sources, following a one-step synthesis recipe developed in previous work [25]. During coevaporation, the substrate temperature was set to an initial temperature of 500 °C, which was increased gradually to 550 °C within the first 10 min. The total time for precursor deposition was 30 min (excluding the cooling step). The AIGSe absorbers were fabricated by coevaporation from Ag, In, Ga and Se sources, following a similar recipe as CGSe. The CGSe and AIGSe absorbers were fabricated in the same coevaporation chamber. The CGSe and AIGSe film compositions were controlled by varying the power applied to each source, based on the flux detected with a mass spectrometer, thereby tuning the evaporation rate. The composition of the films were measured by XRF using a PANalytical Epsilon 5 system.

**Table 1** – Summary of the different fabrication conditions for the chalcogenide top cells and Si bottom cell layers. Any variations from the parameters presented here is explicitly mentioned in the main text.

| Chalcogenide | Top cell | | Bottom cell | | |
|---|---|---|---|---|---|
| | Route | Annealing | n+polySi (nm) | TiN (nm) | p+polySi (nm) |
| **CZTS-D** (Cu$_2$ZnSnS$_4$) | Cosputtering Cu/ZnS/SnS | 575 °C, 45 min, 50 mg S powder | 40, 200, 400 | 2.5, 5, 25 | 40 |
| **CZTS-U** (Cu$_2$ZnSnS$_4$) | Cosputtering CuS/ZnS/SnS | 582 °C, 1 min or 13 min, 80 mg S powder | 200, 400 | 2.5, 5 | 40 |
| **CGSe** (CuGaSe$_2$) | Coevaporation Cu, Ga, Se (one step) | 550 °C 20 min | 200 | 2.5 | 40 |
| **AIGSe** (AgInGaSe$_2$) | Coevaporation Ag, In, Ga, Se (one step) | 550 °C 20 min | 200 | 2.5 | 40 |

Secondary Ion Mass Spectrometry (SIMS) measurements were conducted at several stages of the process, before and after the top cell fabrication, using a Cameca IMS-7f microprobe (both 15 keV Cs$^-$ and 10 keV O$_2^+$ ions used as primary beam). After top cell absorber fabrication, the chalcogenide and TiN layers were chemically etched off in H$_2$O$_2$:4H$_2$SO$_4$ (piranha) and RCA1 solutions, followed by a dilute HF dip. Then, SIMS depth profiles were obtained on the exposed Si bottom cell stack (p+polySi/SiO$_x$/n-Si/SiO$_x$/n+polySi), both from the front side (n+polySi side) and the rear side (p+polySi side). In some SIMS experiments, symmetrically-passivated samples with 40 nm n+polySi on both sides were also included for comparison.

Rutherford Backscattering Spectrometry (RBS) measurements were carried out using the 5-MV NEC-5SDH-2 tandem accelerator at Uppsala University [47], and deploying 2 MeV He$^+$ primary ions on a set of samples with 200 nm n+polySi on both sides, after fabrication of CZTS and CGSe. The RBS solid-state detector ($\approx$ 13 keV of energy resolution in the present chain) was fixed at a scattering angle of 170°. The incidence angle between the sample normal and beam direction was 5°. For each RBS spectrum, the incidence/exit angles were randomized by small steps of ±2° around an equilibrium incident angular position to reduce residual channeling effects on the Si substrate. The measurements were done in the high-dose and low-current regime, aiming to avoid problems with pile-up effects (< 1% for all measurements). Further details on the RBS setup and data analysis can be found elsewhere [48]. Prior to the RBS measurements, the chalcogenide and TiN top layers were etched in Piranha, RCA1 and buffered HF, to eliminate depth effects and measure signal only from the remaining polySi/SiO$_x$/Si stacks.

The current-voltage (J-V) characteristics of the tandem solar cells fabricated in this work were measured at near standard test conditions (STC: 1000 W/m$^2$, AM 1.5G, 25 °C). A Newport class ABA steady-state simulator solar simulator was used. The irradiance was measured with a 2×2 cm$^2$ Mono-Si reference cell from ReRa certified at STC by the Nijmegen PV measurement facility. The temperature was kept at 25 ± 2 °C as measured by a temperature probe on the contact plate. The acquisition was done with 2 ms between points, using four-wire measurement probes. The scan direction was from reverse to forward voltage. The samples were scanned from -1V to 1.5V, to investigate possible non-ideal behaviors across all quadrants (such as S-shape or rollover effects). Some samples were remeasured after 11 months of storage in ambient conditions, using the same conditions but with a new irradiance calibration, to account for long-term variability in the illumination setup. For the external quantum efficiency (EQE) measurements, a QEXL setup from PV Measurements was used. The setup was equipped with a grating monochromator, a power source for adjustable bias voltage and a white light source for light-biasing. The CZTS, CGSe and AIGSe top cells were measured with light-biasing of the Si bottom cell using a high-pass filter at 900 nm, and the Si bottom cell was measured with light biasing of the respective top cells using a band-pass filter from 400 to 500 nm.

To estimate the optical losses in the CZTS/Si tandem devices produced, Monte Carlo ray-tracing optical simulations were done using the SunSolve software by PV lighthouse. The optical constants for the top cell layers were chosen from literature (CZTS [49], CdS, i-ZnO and AZO [50]). The optical constants for the polySi and TiN layers were measured by spectroscopic ellipsometry in reflection mode using an M-2000 rotating compensator ellipsometer from J.A. Woollam.

## 3 Results and Discussion

For a visualization of the structures used in this work, the simplified process workflow and associated characterization of all tandem devices is outlined in **Figure 2**.

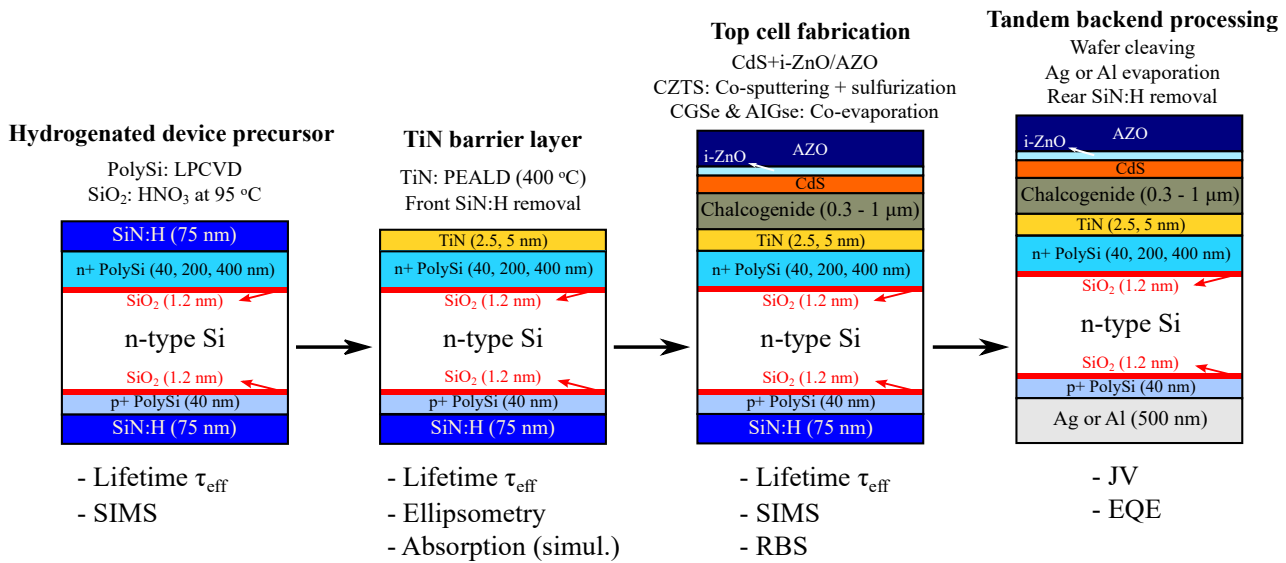

**Figure 2** – Simplified process flow and characterization sequence. From left to right: the TOPCon device precursor wafer, capped with SiN:H on both sides, is followed by the TiN deposition on the n+polySi side. Afterwards, all the top cell layers are fabricated, and finally the rear SiN:H is removed for depositing the metal back contact.

To evaluate the effect of the polySi thickness on the resilience of the Si bottom cell against the top cell fabrication, we used SIMS to measure quantitative Cu profiles in the Si wafer after the synthesis of the top cell absorber layer. As we have reported previously for the case of CZTS, Cu contamination in Si, resulting from the in-diffusion of Cu-containing precursor species during the reactive sulfurization step, is responsible for most of the bottom cell degradation [31,32]. Therefore, the Cu concentration profile in Si is an excellent proxy to study the effects of modifying the Si bottom cell structure. In **Figure 3**, we present a side-by-side comparison of the Cu profiles in Si measured by SIMS, for different TOPCon Si structures. After the CZTS annealing step, the CZTS and TiN layers were etched off, and the measurement was performed on both the front side and the rear side of the wafer. In **Figure 3 (a)**, the Cu concentration profile is shown for a symmetrically passivated wafer with 40 nm n+polySi on both sides (a typical thickness for 1J TOPCon Si cells), and with and without a 25 nm TiN barrier layer. In **Figure 3 (b)**, the same measurement is done on wafers with 200 nm or 400 nm n+polySi on the front, 40 nm p+polySi on the back, and in both cases a 2.5 nm TiN barrier. In all the figures, a Cu accumulation occurs in the polySi regions. The Cu in-diffusion depth profile can be mitigated by roughly one order of magnitude using a thick 25 nm TiN barrier, which we have characterized in previous work [31,32]. However, for 200 and 400 nm of n+polySi, the results show that the Cu concentration decreases by 3 orders of magnitude (99.9%) towards the Si bulk, down to $10^{16}$ cm$^{-3}$, even with a very thin 2.5 nm TiN barrier. In both cases, the rear measurements also reveal a Cu accumulation in the polySi, with a similar Cu concentration compared to the frontside polySi. We can exclude the possibility of this Cu contamination coming directly from the backside (for instance from the substrate holder or susceptor), as the rear-side was systematically protected with a 75-100 nm SiN:H sacrificial layer (see **Figure 2**), or a >25 nm TiN layer (not shown here). Therefore, the backside Cu profile is actually a result of Cu diffusing across the whole Si wafer (350 μm) and accumulating at the back. Since this backside accumulation happens regardless of the frontside n+polySi thickness, we can infer that the Cu distribution is governed by a gettering effect from the polySi layers, rather than a diffusion barrier effect.

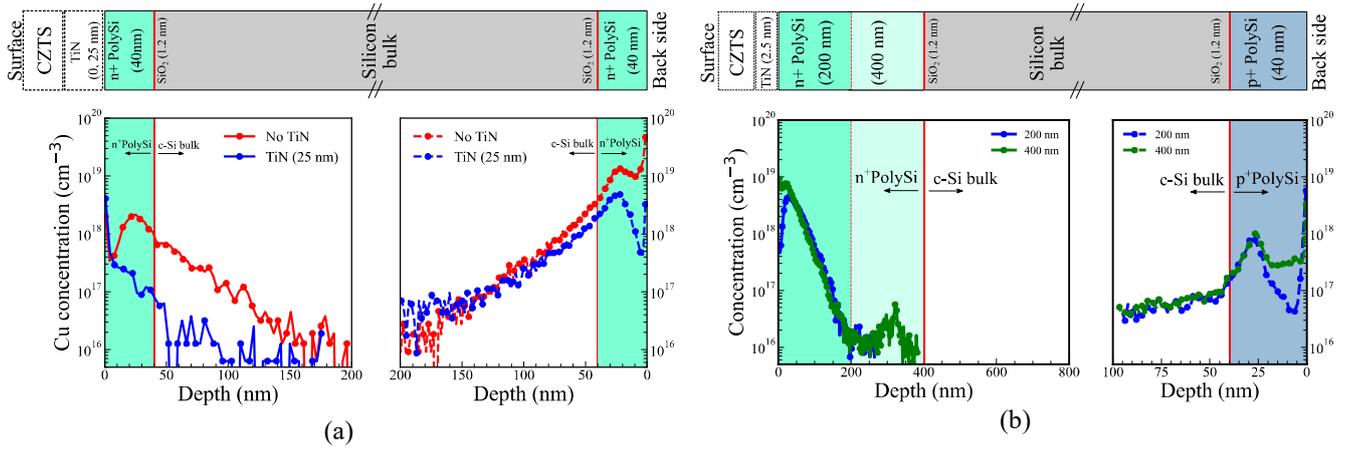

**Figure 3** – Front and rear SIMS profiles of the Cu concentration in Si after CZTS processing for (a) 40 nm n+polySi on both sides, with and without a 25 nm TiN barrier, and for (b) 200 and 400 nm n+polySi on the front side and 40 nm p+polySi on the backside, with a 2.5 nm TiN barrier. The CZTS and TiN barriers were etched off before the measurements.

As a result of this significant reduction of the Cu concentration in the Si bulk, the effective minority carrier lifetime $\tau_{eff}$ of Si after the CZTS sulfurization step was found to be one order of magnitude higher for a 200 nm n+polySi compared to a 40 nm n+polySi, as shown in **Figure 4**. In both cases, a 2.5 nm TiN layer and a 40 nm p+polySi were used. The median lifetime value of 750 µs obtained for the 200 nm n+polySi represents a relatively small degradation from the original lifetime of 1.76 ms of the as-passivated device precursor wafer (before CZTS processing). To clarify the beneficial effect of the increased polySi thickness, it is critical to analyze the effects of the polySi thickness on the passivation quality of the Si device precursor wafer itself, before any top cell fabrication takes place. In the supporting information (SI), in Figure S1 we provide a comparison of the lifetime $\tau_{eff}$ for three device precursor wafers fabricated under similar conditions with 40, 200 and 400 nm n+polySi, both as-passivated and after the SiN:H hydrogenation step. In both cases, the three initial $\tau_{eff}$ were slightly different, albeit within the same order of magnitude. This difference in passivation quality can be attributed to the optimization of the polySi growth conditions as a function of thickness, namely the doping concentration profile, and the annealing profile. For each polySi thickness there is an ideal doping and annealing profile that optimizes the field effect (enhanced carrier selectivity) without excessive dopant diffusion into the c-Si. For that reason, variations of passivation quality with polySi thickness have been frequently reported in literature [39,51,52], and each thickness would require individual optimization. In our internal optimization work, the polySi growth conditions were mainly studied for 40 nm polySi, and the resulting dopant concentration profiles, as measured by SIMS, can be found in Figure S2. Maintaining these conditions, it was observed that the passivation quality improved by increasing the n+polySi thickness to 200 nm. We found this improvement to be systematic and reproducible. As an example, in Figure S3 we studied the situation where a 160 nm polySi was grown directly on top of a symmetrically passivated wafer with 40 nm n+polySi, and the passivation quality improved across several samples, both with and without a SiN:H hydrogenation step. This likely indicates that our polySi growth conditions are slightly more favorable for a 200 nm

n+polySi. Similarly, the thicker polySi layers in this work were done by sequential depositions (40 nm first in all cases, then 160 nm for 200nm, and 200 nm to achieve 400 nm). Therefore, to analyze the bottom cell resilience results presented above and throughout this work, we have to note that the precursor wafer (initial) passivation quality is also changing as a function of the polySi thickness. This change is unavoidable, due to the different optimal parameters required for each polySi thickness. In any case, while the initial lifetime values are within the same order of magnitude (~1-2 ms), the final lifetimes after CZTS fabrication, shown in **Figure 4**, differ by over one order of magnitude. The majority of this effect can thus be attributed to the beneficial protection of the thicker n+polySi layer during the top cell fabrication.

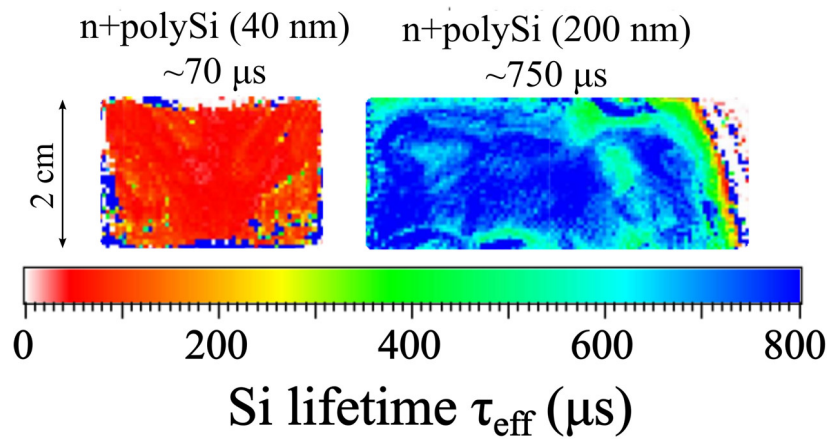

**Figure 4** – Minority carrier lifetime maps of the Si bottom cell after the CZTS annealing step (for CZTS-D samples), comparing the use of 40 nm of n+polySi (left) and 200 nm of n+polySi (right). A 2.5 nm TiN layer and a 40 nm rear p+polySi layer were used in both cases.

This increased bottom cell protection with a thicker n+polySi was also confirmed for samples without any TiN barrier layer. In **Figure 5**, Si lifetime maps of samples with 40 nm n+polySi and 200 nm are compared, for different annealing temperatures used during the CZTS sulfurization step. The resulting lifetimes are again up to one order of magnitude higher when a 200 nm n+polySi is used, and the lifetime is relatively uniform across several cm$^2$ of area near the center of the wafer pieces. Note that the lifetime significantly decreases towards the edges, where the passivation is poorer either due the wafer cleaving steps used to produce smaller chips, or due to edge effects during the different fabrication steps. The results are qualitatively consistent with the SIMS and lifetime results in **Figure 3** and **Figure 4**, respectively, pointing towards the polySi being the largest contributor to the increased protection of the bottom Si cell during CZTS synthesis. However, **Figure 5** also shows that the Si lifetime is quite sensitive to the CZTS annealing temperature, showing that the CZTS synthesis conditions are harsh and near the resilience threshold of this Si configuration. Still, a temperature of 575 °C is well within the typical CZTS sulfurization temperatures used by most groups (550-590 °C), and these Si wafers seem to withstand this temperature with only minor degradation.

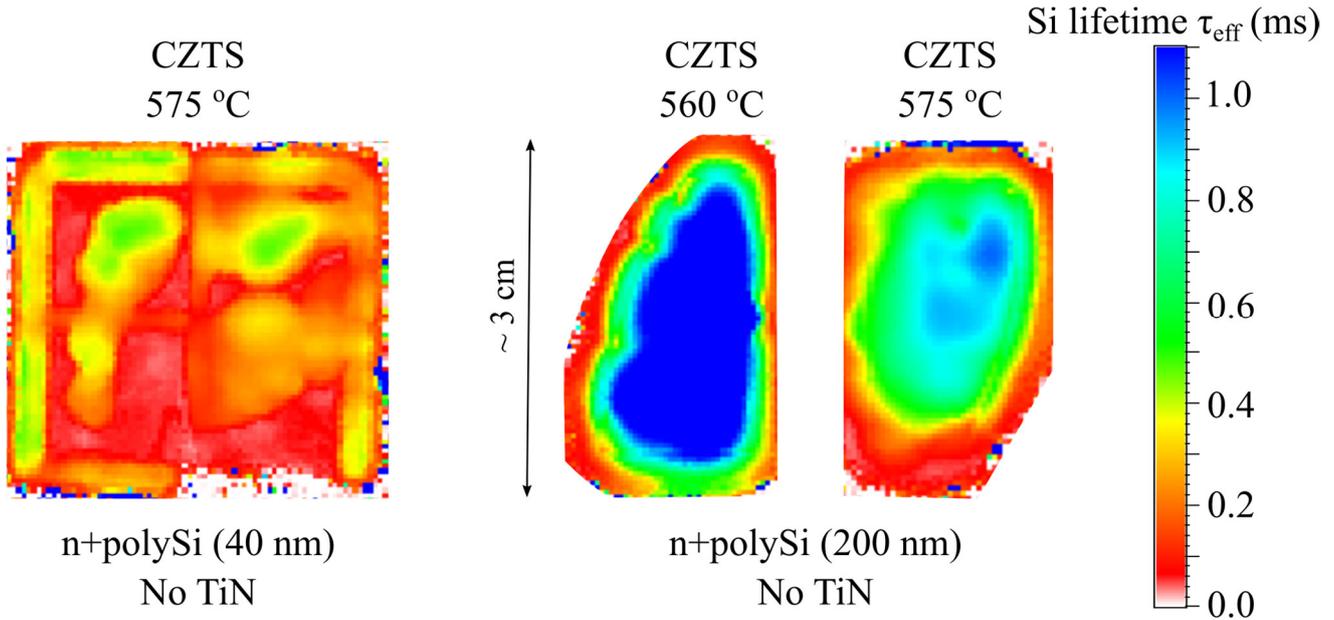

**Figure 5** – Minority carrier lifetime maps of Si after the CZTS annealing step (for CZTS-D samples), comparing n+polySi thickness and CZTS annealing temperature: left, 40 nm n+polySi and 575 °C; center, 200 nm n+polySi and 560 °C; right, 200 nm n+polySi and 575 °C. No TiN layer was used. A 40 nm rear p+polySi layer was used in all cases.

These results also force us to reassess the role of TiN as a "barrier" layer. In fact, our results seem to show that, with this cell architecture, the bottom cell resilience is mainly assured by the increased gettering effect provided by the thicker n+polySi layer, and the resilience is high even without any TiN layer. Therefore, in these conditions, it is not fully adequate to label TiN as a barrier layer. The barrier effect of TiN can indeed be enhanced, for instance by increasing its oxygen content, as we have shown in previous work [32], but does not appear to be significant in our proposed configuration with a thicker polySi. Nevertheless, we have found it useful to keep this TiN layer in the structure for several reasons. First, the adhesion between CZTS and the n+polySi was found to be quite poor, and TiN improved this adhesion issue. Moreover, due to its metallic nature, TiN can improve the conductivity at the interface between the bottom and top cells, and therefore reduce the resistance of the tandem recombination layer. Finally, as mentioned above, TiN can still provide a minor barrier effect that adds additional protection to the Si bottom cell. Based on this, TiN could be more accurately described here as a "buffer" layer rather than a barrier layer. By reducing the TiN thickness down to 2.5 nm, we benefit from these effects and simultaneously maximize the transmittance of the bottom cell structures. In fact, for configurations with a thick n+polySi (200 and 400 nm), we did not find a significant difference in the Cu contamination profiles in Si when comparing TiN layers with thicknesses of 2.5 and 5 nm. The corresponding front and rear SIMS profiles can be seen in Figure S4 (a) and (b), respectively.

As mentioned above, the data suggests that the increased bottom cell resilience is associated with an increase in the gettering of contaminants by the thicker polySi layers. However, the exact type of gettering mechanism taking place and how it can be further controlled remain open questions. Gettering in Si has been studied for decades, and several individual gettering mechanisms have been identified, namely in the work by Myers, Seibt and Schröter [53], which reviews metal-silicide precipitation, segregation into binary phases (e.g. Al-Si or B-Si), atomic trapping by defects, electronic dopant interactions (electrostatic and Fermi level shifting), nonequilibrium processes associated with defect fluxes and, in particular, phosphorous-diffusion gettering, which is known to have a very high gettering efficiency. Similarly, the use of polycrystalline Si as a gettering agent has been long studied [53–55], and the gettering in the specific polySi/SiO$_x$ structures discussed in this work has received considerable attention recently, precisely due to their important role as passivating contacts in high-efficiency Si solar cells [39–41,56]. Although the classification of the different gettering mechanisms outlined above has been a topic of debate [53], they can generally be labelled as either a segregation or a relaxation process. The difference between the two is that segregation gettering results from an equilibrium reached at a given annealing temperature, whereas relaxation results from the non-equilibrium induced in the system during the cooling stage of the annealing [55]. In some cases, the two components can be separated by a forced cooling (quenching), "freezing" the high-temperature distribution of impurities. However, in our case, this is very difficult to achieve due to the high diffusivity of Cu in Si. An estimate of the distance diffused by Cu during the cooling step of the top cell annealing can be obtained from the characteristic diffusion length $\sqrt{Dt}$, which is a scale parameter naturally emerging in most solutions of the diffusion equation. For Cu in Si with a diffusion coefficient $D = 2 \times 10^{-5}$ cm$^2$s$^{-1}$ at 575 °C, even in the best case scenario of a NaOH quench (2000 K/s [57]), resulting in a very low diffusion time $t$ of 0.25 s, Cu would still diffuse through a distance of approximately 22 μm, which is two orders of magnitude higher than the polySi depth range analyzed in our SIMS results. To make matters worse, this "frozen" profile could only be measured immediately after the quenching, because Cu can still diffuse across the entire Si wafer at room temperature in a matter of hours [33]. For that reason, we estimate that the relaxation and segregation components cannot be separated in our gettering process. In any case, an important corollary of this result is that a significant gettering should occur to some degree regardless of the top cell annealing profile, even though it could be further tuned with a slower cooling profile. Indeed, we found similar bottom cell protection levels for the CZTS-D and CZTS-U top cells fabricated in different laboratories with different annealing profiles, as described in the **Experimental Section**. Another notable aspect of this gettering process is that the gettering takes place in a subsequent step (the top cell annealing step), and not immediately after or during the polySi fabrication, where pre-existing contaminants in Si are gettered by the polySi layers. To the best of our understanding, this is a less commonly studied case of polySi gettering, where the contaminants actually diffuse through one of the polySi layers before reaching the Si base, and then are gettered by the front and rear polySi layers. As mentioned above, given the similarity of our results for 200 nm and 400 nm of n+polySi, it is not clear whether this extra "polySi/SiO$_x$ barrier" makes any difference in the bottom cell protection for our current architecture, and this could be studied in future work. In a first approximation, a difference could be expected if the diffusivity of a given contaminant is significantly lower in the polySi/SiO$_x$ stack, compared to Si. In the polySi, the diffusivity of contaminants will generally be worse (*i.e.* higher) than in c-Si due to the prevalence of grain boundary diffusion [58]. This likely explains why the 400 nm polySi case behaves similarly to the 200 nm polySi case, as no direct barrier effect should occur. However, for SiO$_2$, the diffusivity of most transition metals is several orders of magnitude lower compared to that of Si. For Cu at 500 °C, $D_{Si} \sim 10^{-5}$ cm$^2$s$^{-1}$ and $D_{SiO2} \sim 10^{-11}$ cm$^2$s$^{-1}$ [33,59]. Another example is iron, where in the opposite direction it is known that SiO$_x$ reduces the efficiency for the gettering of a pre-existing Fe contamination due to the low diffusivity of Fe in SiO$_2$ (at 1050 °C $D_{Si} = 2.7 \times 10^{-6}$ cm$^2$s$^{-1}$ and $D_{SiO2} = 4.2 \times 10^{-14}$ cm$^2$s$^{-1}$ [55,56]). Moreover, this is also precisely the case enabled by using an effective diffusion barrier such as TiON. Therefore, a general difference between the two gettering cases is that in our case the contamination admitted into the Si base can be mitigated by the diffusion barrier effect, and then regulated by the polySi gettering.

To interpret the effect of the polySi thickness on the gettering efficiency, we find two plausible hypotheses, which are not mutually exclusive. On one hand, the inhomogeneous polycrystalline nature of polySi enables gettering at its defect centers (dislocations, stacking faults) and in particular its grain boundaries, where a high density of defects tend to accumulate [53,55]. The amount of these defect centers depends on the total volume and grain boundary area, and therefore scales non-linearly with polySi thickness. On the other hand, as discussed above, the polySi growth conditions and doping profile change with thickness, and since the gettering is dependent on the resulting polySi properties, there can be an indirect dependence of gettering with polySi thickness [40,55]. Lastly, although not strictly dependent on the polySi thickness, the dopant concentration and the type of dopant are crucial parameters in achieving polySi gettering. For certain growth conditions, intrinsic polySi alone may not provide any significant gettering effect, and the presence of heavy doping is largely responsible for the gettering [40,55]. Notably, gettering has been clearly identified in P and B doping, whereas As and Sb doping do not produce gettering effects [60,61]. In polySi gettering, both heavy P and B doping have been shown to produce significant gettering of transition metals, with P doping displaying the highest gettering efficiency overal [40,41,56]. This is consistent with our results, where there is evidence of Cu gettering taking place in both the n+polySi and the p+polySi layers. We note that despite the complex relationship between the gettering effect and the polySi/SiO$_x$ fabrication conditions, we were already able to demonstrate a significant increase in the bottom cell resilience by simply increasing the polySi thickness, for a given dopant flow during the LPCVD step. So far, we have not directly studied the influence of the doping concentration or the polySi annealing profile on the bottom cell resilience, and therefore we believe there is a significant margin for further improvement in the bottom cell resilience, highlighting the potential of this structure.

In the following, we will present the tandem device results achieved by combining different chalcogenide top cells with this bottom cell structure, and illustrate the improvements achieved by the increased bottom cell resilience. The results for CZTS, produced in two different laboratories (CZTS-D and CZTS-U), are shown in **Figure 6**. A record cell of 6.8% efficiency and $V_{oc}$ 1083 mV was achieved with a 200 nm n+polySi, as shown by the J-V characteristics in **Figure 6 (a)**. The cell was remeasured after being stored for 11 months in air (no encapsulation), and shows nearly the same J-V characteristic, apart from a slightly lower $J_{sc}$ and FF, which we attribute to a known increase in the series resistance in the AZO due to air and humidity exposure [17]. It is known that Si and CZTS devices show good long term stability, but it is interesting to check that the tandem recombination junction and, in particular, the gettered contaminants in the polySi layers, seem to exhibit a similarly stable behavior.

The external quantum efficiency (EQE) of the CZTS/Si tandem devices, shown in **Figure 6 (b)**, corroborates the observation that the bottom cell shows similar resilience to the CZTS fabrication conditions in both laboratories. In **Figure 6 (c),** the boxplots show the device statistics for the best tandem wafer substrate, with a total area of approximately 20 cm$^2$ across several chips.

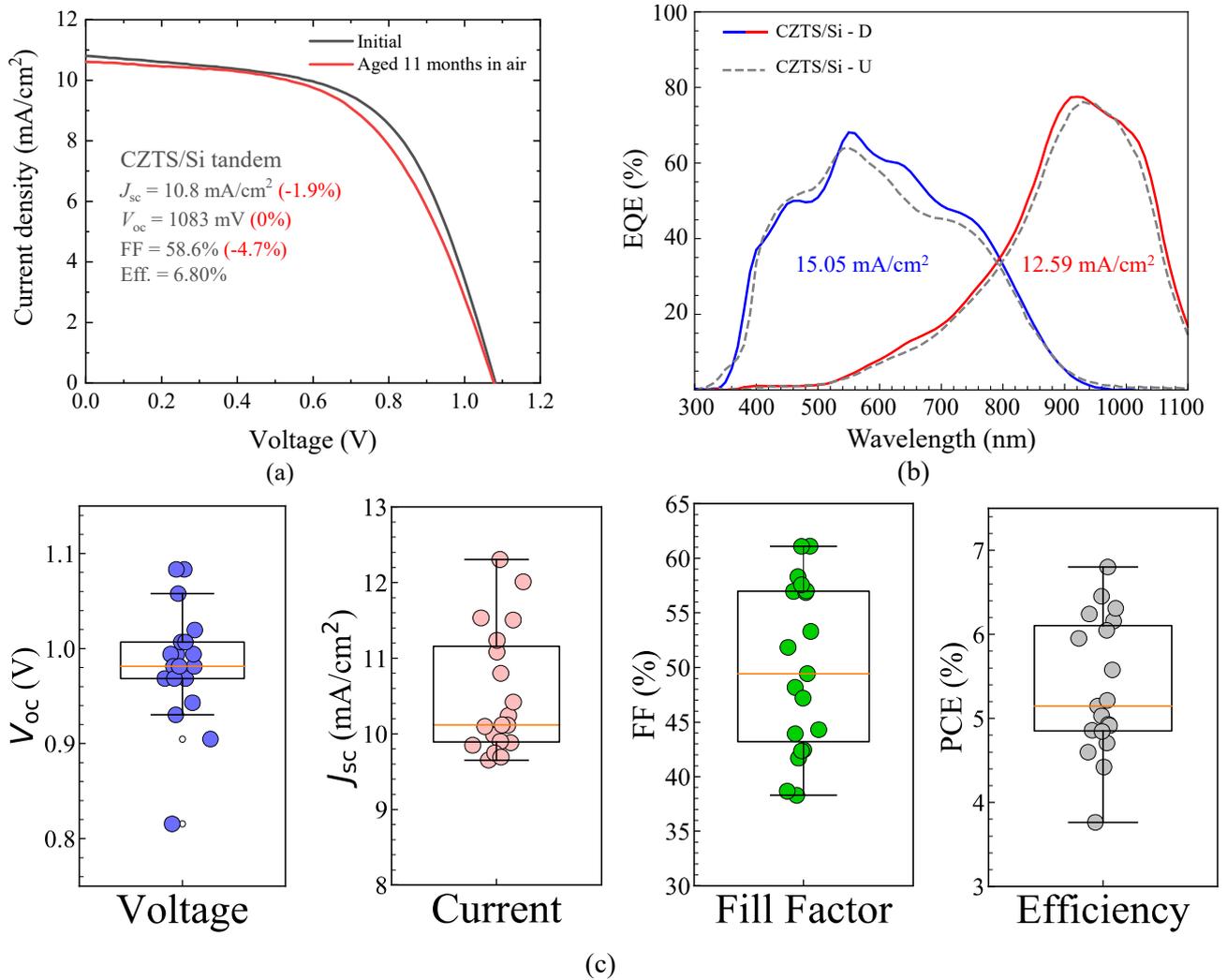

**Figure 6** – PV device results for the 2T CZTS/Si tandem solar cells with 200 nm n+polySi. (a) J-V characteristic of the record CZTS/Si cell before and after 11 months aging in air without encapsulation; (b) EQE of the best CZTS/Si cells with the CZTS prepared in different laboratories; (c) Boxplot statistics for the best CZTS/Si tandem chip (20 cm$^2$).

Although there is some variability in the results, it should be noted that both the top and the bottom cells can contribute to this variability. Indeed, uniformity in chalcogenide materials is challenging due to the possibility of compositional variations across large areas, especially in smaller-scale research setups. In particular, in the CZTS-D sputtering process, across a total processing area of around 20 cm$^2$ it is not uncommon to find a standard deviation corresponding to 25% of the mean values of all J-V parameters [62]. On the other hand, the Si substrates are processed from 100 mm diameter wafers in cleanroom conditions, and show a uniform carrier lifetime even after the top cell fabrication. In **Figure S5**, we present further carrier lifetime mappings of Si for different CZTS-U annealing conditions, showing that when a 200 nm n+polySi is used, uniform Si lifetimes above 0.5 ms can also be obtained. Therefore, we attribute a significant part of this variability to the CZTS top cell itself.

Analyzing the Si bottom cell EQE response in greater detail, in **Figure 7 (a)**, shows the clear protection effect of the thicker n+polySi layers. In previous work [32], we have introduced a simple mathematical model to explain these results based on the relationship between the collection efficiency in Si and the effective carrier lifetime in Si $\tau_{eff}$, which is determined by the surface recombination velocity at the Si interfaces and by the bulk lifetime. Our EQE results represent a significant improvement over that previous work, and are consistent with the improvement in $\tau_{eff}$ by one order of magnitude. While the difference between 40 nm and 200 nm of polySi is significant, the EQE response from the samples with 400 nm of n+polySi is only slightly higher than for 200 nm n+polySi, which is consistent with our SIMS results, showing a very similar gettering effect in both cases. For that reason, any thickness in this range should be suitable to achieve a good bottom cell resilience. Regarding the effect of the TiN barrier thickness, shown in **Figure 7 (b)**, we find that 2.5 nm achieves the best bottom cell response, which is an indicator of the improvement in the trade-off between transmittance and bottom cell protection, as mentioned above.

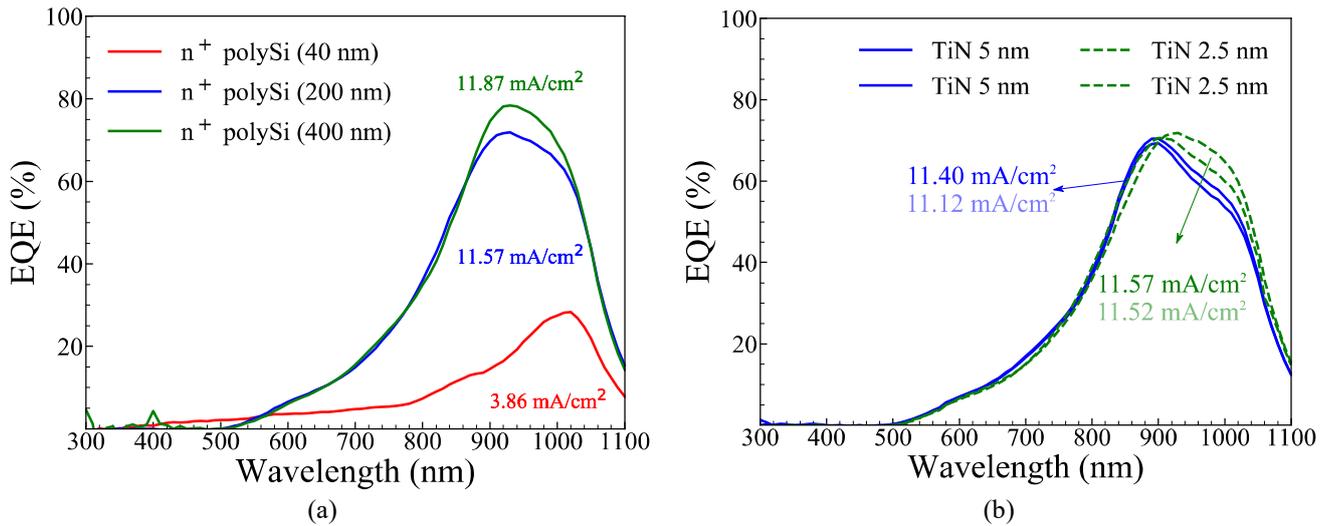

**Figure 7** – EQE comparisons for CZTS/Si tandem cells showing only the Si bottom cell response for (a) different polySi thicknesses and 2.5 nm TiN, and (b) different TiN thicknesses, with an n+polySi thickness of 200 nm.

To provide a more detailed analysis, we compare our results to a Monte Carlo ray-tracing optical simulation, which quantifies the optical losses in all the layers of our tandem structure, assuming flat surfaces, as described in the **Experimental Section**. The inputs and results of this optical simulation can be found in the SI, in **Figure S6**, **Figure S7** and **Table S1**. The simulation allows us to obtain a rough upper limit estimate of the maximum current extracted by the Si bottom cell, assuming ideal operation and only optical losses. For a 200 nm n+polySi, the model predicts a maximum $J_{sc}$ of 12.0 mA/cm², which is very close to our experimental value. This means that the collection efficiency in Si is nearly ideal, which is consistent with the higher $\tau_{eff}$ values resulting from the increased bottom cell resilience. The overall low Si $J_{sc}$ predicted by the model is mostly a result of large surface reflection losses (no texturing or antireflection coatings were used), and an excessive absorption on the CZTS top cell (poor bandgap matching with Si and sub-bandgap absorption).

The same Si bottom cell structure, with 200 nm n+polySi and 2.5 nm TiN, was also used to fabricate CGSe/Si and AIGSe/Si tandem devices. For the CGSe/Si tandem, an efficiency of 5.1% and a $V_{oc}$ of 1.3V was achieved, as shown in the J-V characteristic of **Figure 8 (a)**. The corresponding EQE from this cell, shown in **Figure 8 (b)**, reveals a poorer EQE response from both subcells, compared to the previous CZTS/Si case. To interpret this result, we traced back the location of the chip corresponding to this cell, and we identified that the chip was located in a corner region of the Si wafer, with a significantly lower lifetime, as shown in **Figure 8 (c)**. As we explained previously, it is common for our Si wafers to have poorly passivated edges, as no wafer edge isolation methods were implemented in this work. Therefore, this edge piece is not representative of the behavior of the bottom Si substrate, which is illustrated by its significantly lower lifetime ($\tau_{eff}$ < 200 μs) compared to regions in the wafer interior, where $\tau_{eff}$ > 600 μs despite the high temperatures during the coevaporation of CGSe. However, the CGSe top cell achieved the best conditions in this region, which resulted in the highest tandem efficiency overall for this batch. For comparison, we include in **Figure 8 (d)** the EQE results for two chips in different regions of the Si wafer, away from the edges. **Figure 8 (d)** reveals that the EQE signal from the Si bottom cell is clearly higher, which is associated with the higher $\tau_{eff}$ in this central region of the wafer. However, for both CGSe batches developed in this work, with thicknesses of 500 nm and 1 μm, the performance of the top cell was poorer compared to that in the edge region. Still, due to the improved performance of the bottom cell, the resulting tandem efficiency reached a maximum of 5.5%, although with voltages below 1.1 V, substantially lower than the best edge piece.

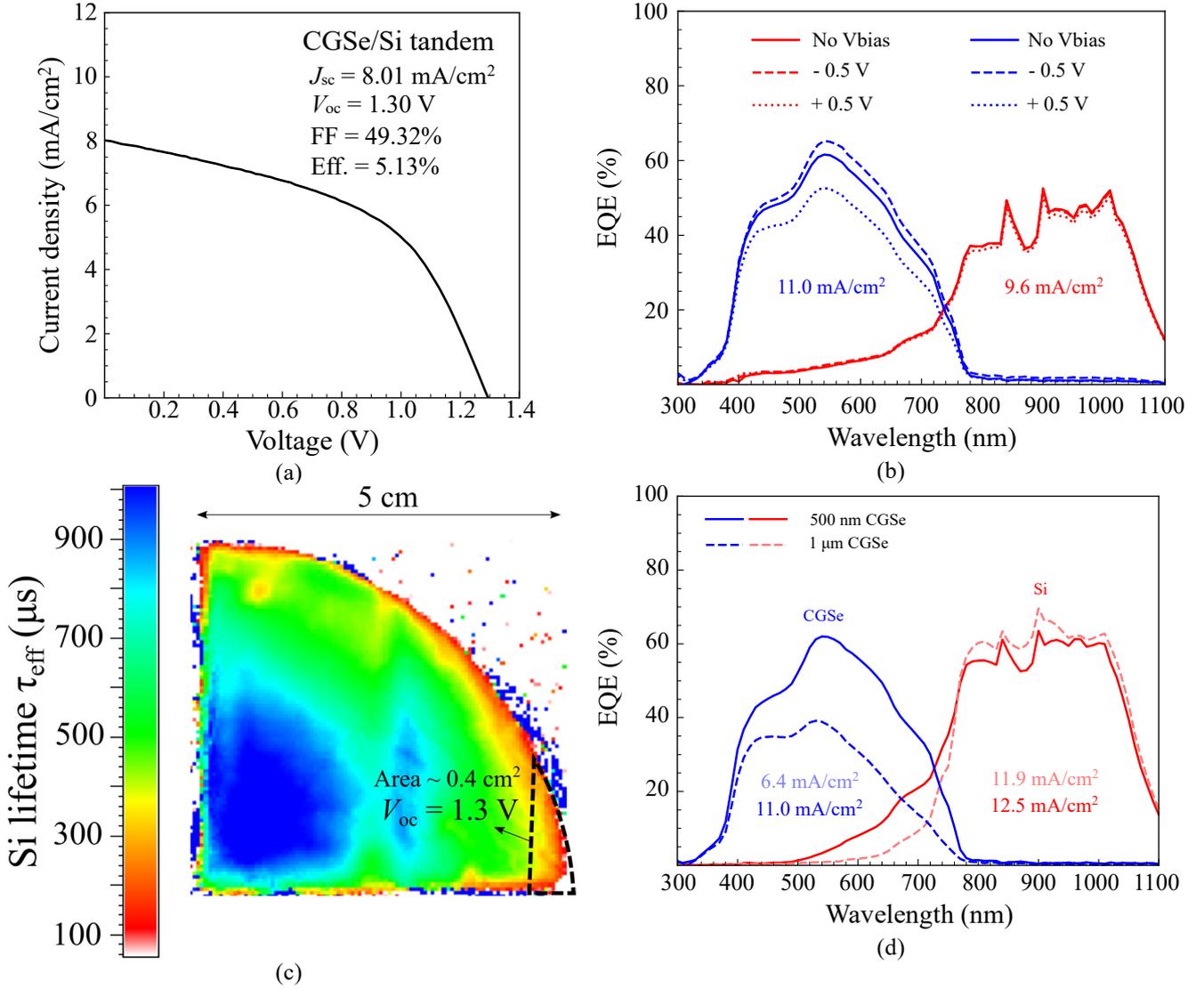

**Figure 8** – PV device results for the CGSe/Si tandem integration. (a) J-V characteristic of the CGSe/Si tandem cell with the highest $V_{oc}$. (b) Corresponding EQE of the highest $V_{oc}$ cell. (c) Si lifetime map after the fabrication of 500 nm of CGSe, showing the location and area of the highest $V_{oc}$ cell. (d) EQE of the highest efficiency CGSe/Si tandem cell, from an inner wafer region, showing an improved Si bottom cell response.

The overall device statistics are shown in the boxplots of **Figure 9**, where a large variability of the results can be confirmed. As mentioned before, large-area reproducibility can be a major challenge in chalcogenide materials. However, although we did not succeed at this stage to match the best quality CGSe top cell with a standard high-lifetime Si bottom cell, we note that our record device already features nearly the same $V_{oc}$ as the current best CGSe/Si tandem achieved thus far, with 9.7% efficiency and a $V_{oc}$ of 1.33 V [36], despite having almost half the efficiency. To understand the meaning of this comparison, we have measured the bottom cell $V_{oc}$ by partially scribing off the top CGSe cell in a corner chip. Based on this measurement, we estimated a $V_{oc}$ > 575 mV for Si and $V_{oc}$ < 750 mV for CGSe across the wafer. By comparison, the record CGSe/Si device exhibitted 774 mV and 558 mV, respectively. Therefore, our Si bottom cell is outputting a higher voltage even in corner locations with reduced lifetime. This is the fundamental advantage of using a high-efficiency passivated contact structure such as TOPCon

Si. In this structure, a $V_{oc}$ exceeding 700 mV is possible when the lifetime is near 1 ms, which this resilient architecture allows. Indeed, even though we have only achieved an external $V_{oc}$ up to 685 mV based on 1J Si backend optimization (not shown here), we have already achieved 660 mV for Si bottom cells in well-protected tandem devices (more details can be found in ref. [32]), highlighting the potential of this bottom cell structure.

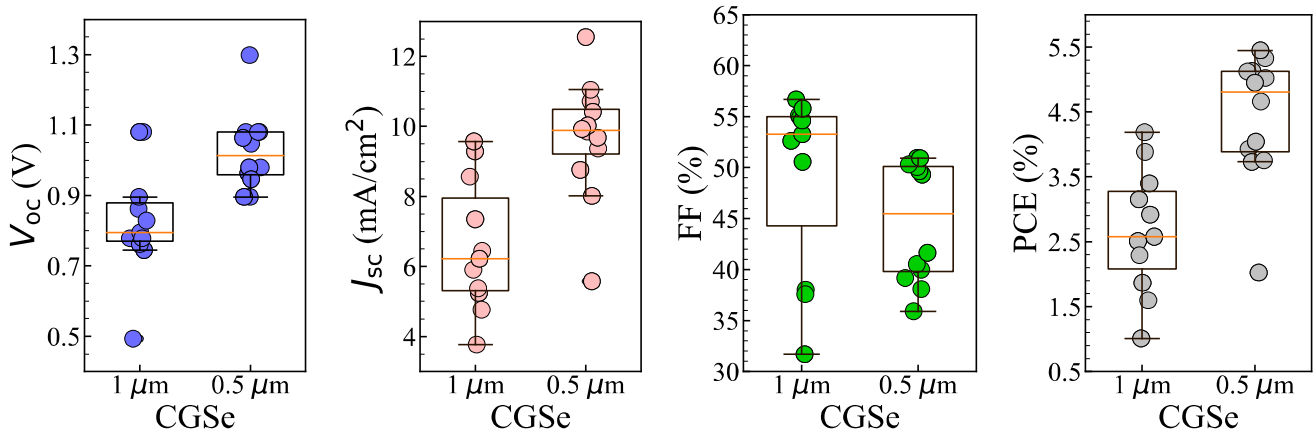

**Figure 9** – Boxplot statistics for the best CGSe/Si tandem chip, from a total area of 20 cm$^2$.

Lastly, the results for the AIGSe/Si tandem are shown in **Figure 10**, in comparison to the CGSe/Si results. The Si lifetime was found to be uniformly higher in the AIGSe case, as shown by the lifetime maps in **Figure 10 (a)**. Both wafer quarters belong to the same Si precursor wafer, and therefore have a comparable initial Si lifetime. On the top cell side, the fabrication conditions (annealing time, temperature and Se atmosphere) were identical, and the only processing difference is that the Cu source was replaced with Ag and that an In source was added in the coevaporation step. Therefore, it is likely that the higher $\tau_{eff}$ for the AIGSe case is related to the absence of Cu diffusion into the bottom cell structures. The EQE results, in **Figure 10 (b)**, show that the AIGSe/Si tandem achieves a comparatively higher EQE on the Si side, with an implied $J_{sc}$ of up to 14.5 mA/cm$^2$. This is a result of both a higher Si $\tau_{eff}$ and a higher bandgap of AIGSe, compared to the CGSe case. However, despite the promising EQE results, we were not yet able to demonstrate corresponding high efficiencies for AIGSe/Si tandems. Of the two AIGSe/Si batches produced in this work, in one case the AIGSe top cell exhibitted a very low shunt resistance, and only the bottom cell response could be measured. In the other case, we saw very low tandem voltages and fill factors, indicating that the top cell could be exhibitting insulating behavior, secondary phases or carrier inversion (from the expected p-type to n-type), which can occur for very off-stoichiometric AIGSe films [63,64]. We note that this is the first attempt at developing an AIGSe/Si tandem, and further optimization will be required. All details on the J-V characteristics and associated boxplot statistics can be consulted in the SI, in **Figure S8** and **Figure S9**. Using XRF, we were able to confirm that both the CGSe and AIGSe films deviated slightly from their ideal composition region, as shown in **Table S2**.

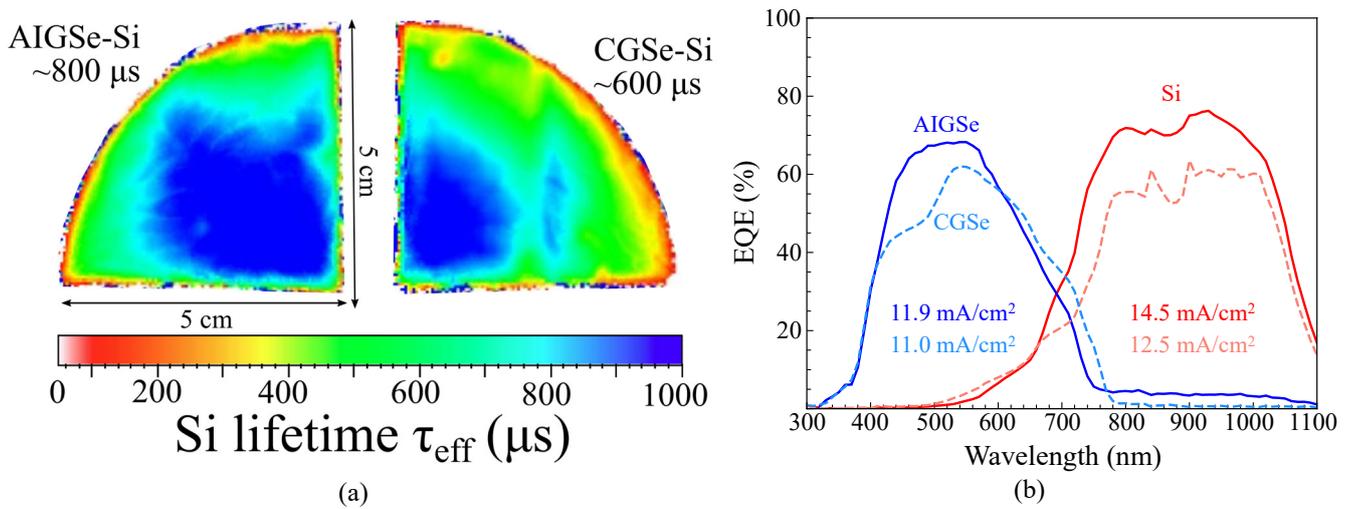

**Figure 10** – (a) Si lifetime median values and maps comparing the impact of AIGSe (left) and CGSe (right) top cell processing. (b) EQE comparison for the best AIGSe/Si and CGSe/Si tandem cells.

Nevertheless, although the full potential of these CGSe/Si and AIGSe/Si tandems could not be demonstrated yet, this work shows that the Si bottom cell can successfully withstand the processing steps of these chalcogenide top cells with minimal performance deterioration. More concretely, we found a clear difference in the bottom cell degradation depending on the top cell. Starting with a series of Si precursor wafers from the same batch, with initial lifetimes close to 1 ms, we consistently found a trend of AIGSe > CGSe > CZTS for the Si lifetime after the respective top cell processing, as shown in **Figure 11**.

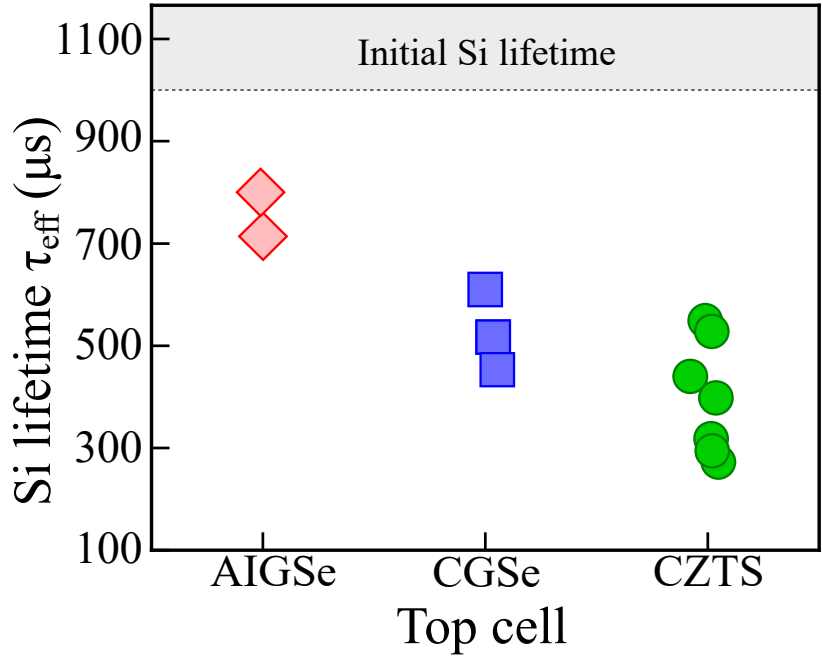

**Figure 11** – Comparison of the impact of fabricating different top cell materials on the Si bottom cell. Each point represents a batch with a minimum of 20 cm$^2$ of area.

As discussed throughout the text, the difference in resulting lifetime is related to specific aspects of the fabrication of these top cell materials. The difference between CGSe and CZTS is likely related to the lower processing temperature (550 °C vs 575-580°, respectively), as we have seen a similar effect when the CZTS annealing temperature is reduced (see **Figure 5**). Not only temperature, but we found the annealing time to be also critical for the bottom cell deterioration. By measuring SIMS at the front and rear for different annealing times, we find a substantially lower Cu profile into the Si bulk when the CZTS-U annealing time is reduced from 13 min to 1 min (the details can be consulted in **Figure S5**). In the case of AIGSe, the processing temperature is similar to CGSe, but there is the additional benefit of not having Cu in the top cell material, which is the main source of contamination diffusing into Si in Cu-based chalcogenides. To confirm this hypothesis, we measured SIMS after the AIGSe and CGSe top cell processing. In **Figure 12 (a)**, the Cu concentration depth profiles originating from the CZTS, CGSe and AIGSe fabrication are compared (red lines), along with the Ag concentration from AIGSe (gray line). The Cu profile is more pronounced for CZTS, although the surface Cu concentration for CGSe is higher. As a result, the corresponding Si lifetimes are similar, with the CGSe case being slightly higher on average. Interestingly, a Cu profile is clearly detected from AIGSe, even though no Cu was intentionally introduced during the AIGSe fabrication. Since AIGSe and CGSe are fabricated in the same coevaporation chamber, the Cu source here could arise from cross-contamination (for instance from the sample holder) as the temperature increases to >500 °C during the deposition. We have double-checked this effect by comparing the depth profile for the $^{63}$Cu and $^{65}$Cu isotope ratio. Whereas in the CZTS and CGSe cases the Cu isotope ratio matched the natural abundances, there was a small deviation for the AIGSe sample, indicating that the Cu concentration may be overestimated due to mass interferences (for instance with Na species). Note that this Cu concentration profile is one order of magnitude lower than the CZTS and CGSe cases, and is close to the SIMS detection limit (~ $6 \times 10^{15}$ cm$^{-3}$ for Cu in Si). In any case, this possible Cu contamination could explain why even in the AIGSe case there is a slight decrease in lifetime compared to the initial precursor wafer lifetime.

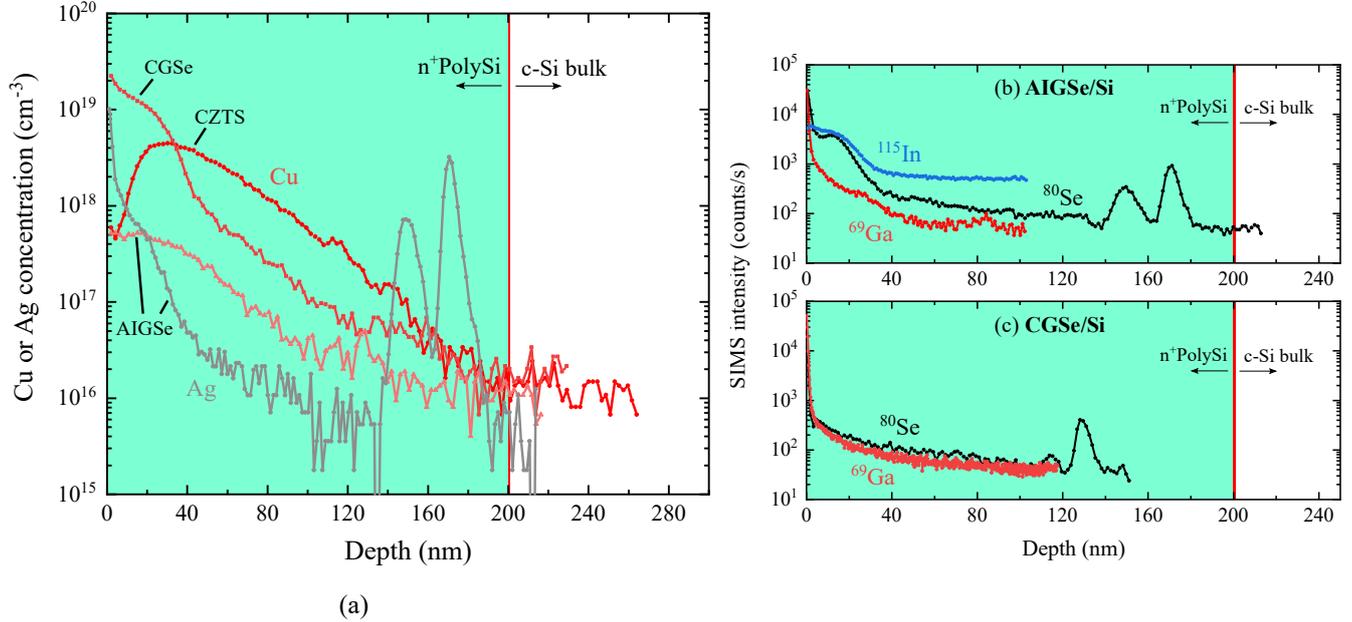

(a)

**Figure 12** – (a) SIMS measurements comparing the Cu concentration profile in Si after fabrication of CGSe, CZTS and AIGSe (red-colored lines), and Ag concentration profile from AIGSe (gray line). (b)-(c) Qualitative SIMS profiles of the elements in AIGSe and CGSe, respectively.

This result, although unexpected, illustrates the general contamination problem of processing Si wafers in multi-purpose or "dirty" chambers. The higher the temperature is, the harder it is to avoid cross-contamination, and this will be a general obstacle in achieving monolithic Si-based tandems, especially with transition metal contaminants. By including gettering-enabling structures in the Si bottom cell architecture, we are able to mitigate this problem and increase the tolerance to bottom cell contamination. Analyzing the other contaminating elements, in **Figure 12 (b)**, shows a similar 3 order of magnitude decrease in the signals of In, Ga and Se in the thick n+polySi, with a nearly flat profile into the Si bulk. Based on similar measurements for CZTS [31], we do not expect these elements to be as problematic as Cu, although we cannot resolve the contribution of each contaminant to the measured Si lifetimes. Interestingly, both the Ag and the Se profiles exhibit local peaks near the polySi/Si interface. This profile could be related to the sequential polySi deposition (40+160+200 nm), which could locally change the SIMS matrix effects or create an additional physical interface where contaminants accumulate. Note that the peak in the 400 nm n+polySi curve for the CZTS/Si case in **Figure 3 (b)** and **Figure S4 (a)** could be a similar effect. The origin of this effect requires further investigation, although it should not affect the general conclusions of this work.

In order to further study the benefical effects of a thicker 200 nm polySi, we have prepared a CZTS and CGSe series on Si for RBS analysis at the front and rear of the Si wafer. As with SIMS, the chalcogenide absorber and TiN were etched in piranha, RCA1 and HF. However, in this case, the Si bottom cell structure is different, having a symmetrical passivation with 200 nm n+polySi on both sides. This way, by doing a comparative RBS and lifetime mapping study, we are able to simultaneously compare the effects of having a thicker polySi on both sides, and search for quantitative measurements of any other possible contaminants such as S, Se, Sn or Ga, which are all heavier than Si and therefore can be well-resolved with RBS (except Zn, which overlaps with Cu). The results of the study are displayed in **Figure 13**. For the CZTS case, the Si lifetime is mapped for two different annealing times (1 min and 13 min, with the CZTS-U process) and two different TiN thicknesses (2.5 nm and 5 nm), as shown in **Figure 13 (a)**. For the CGSe case, TiN thicknesses of 2.5 and 5 nm are compared, and the effect of a NaF post-deposition treatment (PDT) on the CGSe absorber is studied, as depicted in **Figure 13 (b)**. Note that here the initial Si lifetime is near 6 ms for both cases, which is related to a superior passivation quality by the n+polySi used on both sides, compared to n+polySi and p+polySi for tandem device wafers, as we have discussed before [31]. It is again evident that a larger lifetime degradation is occuring in the CZTS case. However, a notable result is that the inner region of the wafer chips always displays lifetime values of at least 2 ms, supporting the previous results of a high bottom cell resilience when a thicker polySi is used. For the CGSe case, the inner region of the wafers shows nearly no degradation in lifetime at all, suggesting that the bottom cell was completely protected in this case. Additionally, the NaF PDT shows no influence on the bottom cell, which opens up possibilities for further optimization of the chalcogenide top cells. We note that this NaF PDT treatment was attempted as part of the optimization work for the CZTS, CGSe and AIGSe top cells, but due to the focus on the bottom cell in this work, these results are not presented here for the sake of brevity. The RBS results, displayed in **Figure 13 (c)-(f)**, show the common feature of no contaminants found within the present detection limit (discussed below), other than a small presence of Ti, which is likely related to some residual TiN not completely removed by the wet etching process, a feature which we have noted before [31]. Due to the lack of signal from all the above-mentioned elements, we can safely establish upper limits for their concentration in Si. Consistent with previous SIMS measurements [31], we expect these concentrations to be close to the detection limits of SIMS, and have minor impact on the Si bottom cell. In the particular case of Cu, the most problematic contaminant, the detection limit in our present RBS measurements is found to be $\approx$ 200 ppm in the Si bulk, which corresponds to $10^{19}$ cm$^{-3}$ (and $\approx 5 \times 10^{14}$ at/cm$^2$ on the Si surface). This Cu concentration upper limit is in good agreement with all our SIMS measurements (**Figure 3**, **Figure 12 (a)**, **Figure S4** and **Figure S5**), where the Cu concentration is always well below $10^{19}$ cm$^{-3}$, and only approaches this value close to the polySi surface, away from the Si bulk. Into the Si bulk, the Cu concentration appears to fall below $10^{16}$ cm$^{-3}$, which is still slightly higher than its solubility at the top cell annealing temperatures, $\sim 10^{15}$ cm$^{-3}$ [65]. On the other hand, it has been well-established that in these transition metals only a fraction of $10^{-3}$-$10^{-4}$ of their solubility in Si results in electrically active (*i.e.* detrimental) defects [66], and indeed Cu precipitates in Si can deteriorate the carrier lifetime at concentrations as low as $10^{12}$-$10^{13}$ cm$^{-3}$ [33]. Therefore, in our case, even though some Cu contamination may reach the Si bulk, the gettering effect seems to maintain this concentration at a level just low enough to still obtain highly efficient Si devices. This, in turn, would explain why the Si lifetime results are quite sensitive to the top cell annealing temperature, as we are near the resilience threshold of this structure, especially in the case of CZTS. As for the other contaminating elements (S, Se, Sn and Ga), assuming a reasonable 100 ppm detection limit for RBS implies that the concentration of other contaminants should not exceed $5 \times 10^{18}$ cm$^{-3}$. However, we note that the RBS sensitivity depends on the atomic mass of the elements and specific instrument aspects (such the charge collection of the backscattering spectra) so this value should be taken as a rough upper limit estimate. Given that the SIMS intensity for In, Ga and Se decreases by 3 orders of magnitude (99.9%) in the 200 nm n+polySi (see **Figure 12 (b)**), the actual bulk Si contamination could be lower by a similar factor.

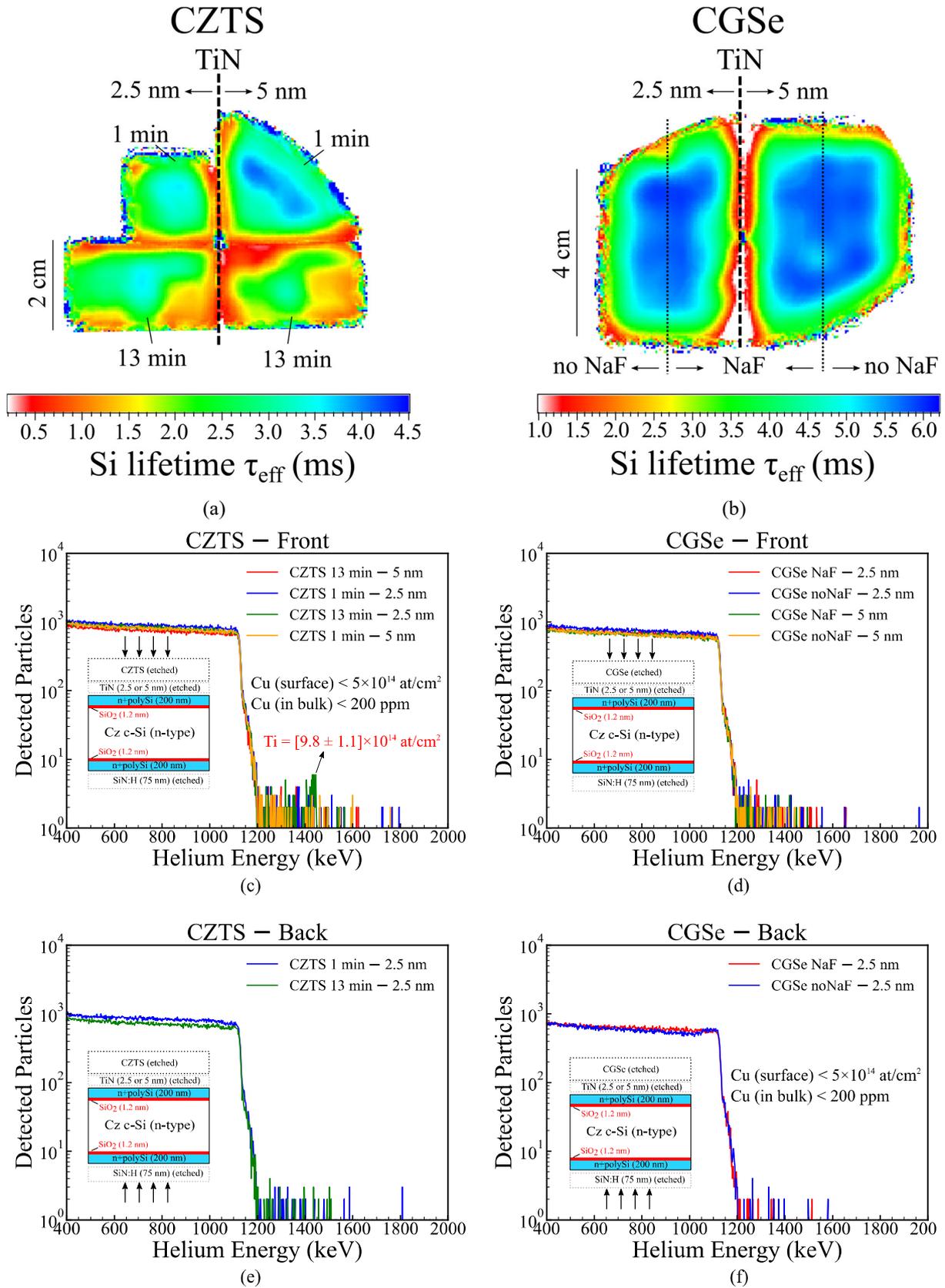

**Figure 13** – Comparative RBS analysis of a CZTS/Si and CGSe/Si batch on symmetrically-passivated wafers with 200 nm n+polySi on both sides. (a) Si lifetime maps comparing different CZTS annealing times and TiN thicknesses. (b) Si lifetime maps for CGSe processing, comparing different TiN thicknesses and a NaF post-deposition treatment. (c)-(f) show RBS

results for the Si wafers after CZTS or CGSe and TiN etching, from the front or rear sides of the wafer, for each fabrication condition as shown in the insets.

Considering these results, the fact that our best tandem device was achieved with CZTS on Si is surprising, given that, of the three compounds, CZTS has the worst bandgap matching with Si and has the harshest fabrication conditions. However, we note that the largest volume of experimental batches and optimization in this work were done on the CZTS/Si pair, and further work is required to find the most favorable conditions to fabricate CGSe and AIGSe on Si substrates. Nevertheless, this comparative work was successful in that it allowed us to establish our Si bottom cell architecture as viable for future work on high efficiency chalcogenide/Si tandems. Based on our results, we can construct some basic estimates on the potential improvements in these tandems. In the short term, just assuming very conservative values such as a $V_{oc}$ of 750 mV for the top cell and 600 mV for the bottom cell, along with a tandem $J_{sc}$ of 12 mA/cm$^2$ and FF of 65%, would result in a tandem efficiency of 10.5%. This estimate is only based on typical values for the top cells, and a modest voltage for the bottom cell which we have already demonstrated in our tandem work. A medium-term estimate can be obtained by taking the state-of-the-art values for both cells. Considering the record CGSe device with 1.0V [25], a bottom cell $V_{oc}$ of 660 mV as achieved in our work, a tandem $J_{sc}$ of 16 mA/cm$^2$ and a FF of 70% would result in a tandem efficiency of 18.6%. This value assumes only that already demonstrated state-of-the-art could be combined in a tandem, and no further developments on either subcell are assumed. Given the results of our work, showing the compatibility of Si with the synthesis of these different chalcogenide materials, this could be a realistic possibility. For a long-term scenario estimate, if we assume that the bandgap-voltage offset (also called $V_{oc}$ deficit) can be reduced in these wide-bandgap chalcogenide materials, down to a value around 0.5 V, therefore assuming a top cell $V_{oc}$ of 1.2 V, a bottom cell $V_{oc}$ above 700 mV (enabled by $\tau_{eff}$ > 1 ms), a tandem $J_{sc}$ of 19 mA/cm$^2$ and a FF of 80%, would result in a tandem with 29% efficiency. In order to achieve this value, further advances in the efficiency of wide-bandgap chalcogenides will be required. Alternatively, other classes of photovoltaic materials could emerge in the future, that may be combined with this Si bottom cell architecture due to its general resilience to high-temperature processes and transition metal contamination. In fact, this is not only a possibility in photovoltaics, but also in photoelectrochemistry such as water splitting, where tandem configurations are also desirable due to the high photovoltages required to drive the electrochemical reactions. Due to its general resilience, this work opens up the possibility of using this Si bottom cell architecture in combination with several known photocathode and photoanode materials for a monolithic tandem. Since the gettering effect can be leveraged to some degree both with n+polySi and with p+polySi layers, this means that our Si bottom cell structure can be easily used with both n-type and p-type top cell materials, simply by flipping the Si device wafer. We will be exploring these possibilities in future work.

## 4 Conclusion

In this work, a comparative study of different chalcogenide-on-Si tandem cells was carried out, by fabricating CZTS, CGSe and AIGSe on Si across different laboratories. The effect of changing the thickness of the polySi/SiO$_x$ passivating contacts was studied in detail, and a significant improvement in the resilience of the Si bottom cell was found generally across all different top cells, during their high-temperature fabrication. By increasing the polySi thickness from the nominal 40 nm to 200 and 400 nm, a one order magnitude improvement in the Si effective minority carrier lifetime was demonstrated. This improvement was correlated with a significantly higher accumulation of impurities (gettering) in the thicker polySi regions, leading to a decrease in 3 orders of magnitude (99.9%) in the concentration of impurities diffusing into the c-Si bulk. Compared to the 40 nm polySi, the 200 and 400 nm polySi layers effectively gettered the contaminants originating from the top cell during their high-temperature fabrication, confining the impurities within the polySi and therefore away from the interfaces and active region of the Si bottom cell. The gettering effect includes both segregation and relaxation gettering components, and is correlated with the heavy B and P doping and polycrystalline nature of the polySi layers, determined during the polySi growth conditions.

A trend in decreasing Si lifetime with the order AIGSe (least Si degradation), CGSe and CZTS (harshest Si degradation) was found to be correlated to the quantitative Cu depth profile from the associated contamination of Si during the respective top cell synthesis. Nevertheless, in all cases, by using 200 nm and 400 nm of polySi, the Cu concentration is effectively kept below $10^{16}$ cm$^{-3}$, leading to effective Si lifetimes above 500 μs across large areas (up to 20 cm$^2$), compared to < 50 μs in the previously-reported configuration with 40 nm polySi. As a result, the high-efficiency polySi/SiO$_x$ passivating contacts retain their excellent properties after the top cell synthesis, leading to a current collection efficiency and voltage output for the Si bottom cell comparable to the ones obtained in similar single-junction Si devices. Moreover, this improvement was achieved despite a simultaneous decrease in the thickness of the TiN interfacial layer down from 25 to just 2.5 nm, thereby improving the transmittance to the bottom cell. The constraint of the TiN layer being needed as effective diffusion barrier layer is therefore relaxed, as most of the protection effect is ensured by the polySi gettering effect. The TiN can still be kept as a buffer layer to improve the electrical interconnection and adhesion between the subcells, but there is now an additional possibility to use alternative interfacial layers. Based on this result, this work presents a significant advance in the resolution of the trade-off triangle limiting the compatibility of chalcogenides on Si, and potentially equivalent inorganic semiconductors fabricated under harsh conditions (high-temperatures and/or transition metal contaminants). By enabling a high Si bottom cell lifetime of ~1 ms, we estimate that the bottom Si cell architecture proposed in this work may enable chalcogenide/Si tandem cells to surpass 20%+ efficiency even without assuming any further improvements in the state of the art of both materials, and may approach 30% if the efficiency of wide-bandgap chalcogenides further improves.

**Acknowledgments**

A. Assar and F. Martinho contributed equally to this work. This work was supported by a grant from the Innovation Fund Denmark (Grant 6154-00008A), by two grants from the Swedish Foundation for Strategic Research (grants RMA150030 and SSF-RIF14-0053), and by the Swedish Research Council (VR, 2019-04793, and contracts # 821-2012-5144 and # 2017-00646-9 for the operation of the Tandem accelerator at Uppsala University).